**Anatomical-Guided Attention Enhances Unsupervised PET Image Denoising Performance**

Yuya Onishi[1*], Fumio Hashimoto[1], Kibo Ote[1], Hiroyuki Ohba[1], Ryosuke Ota[1], Etsuji Yoshikawa[1], Yasuomi Ouchi[2]

[1] Central Research Laboratory, Hamamatsu Photonics K. K., 5000 Hirakuchi, Hamakita-ku, Hamamatsu 434-8601, Japan.

[2] Department of Biofunctional Imaging, Preeminent Medical Photonics Education & Research Center, Hamamatsu University School of Medicine, 1-20-1 Handayama, Higashi-ku, Hamamatsu 431-3192, Japan.

* Corresponding author: yuya.onishi@hpk.co.jp

Yuya Onishi and Fumio Hashimoto contributed equally to this work.

**Abstract**

Although supervised convolutional neural networks (CNNs) often outperform conventional alternatives for denoising positron emission tomography (PET) images, they require many low- and high-quality reference PET image pairs. Herein, we propose an unsupervised 3D PET image denoising method based on an anatomical information-guided attention mechanism. The proposed magnetic resonance-guided deep decoder (MR-GDD) utilizes the spatial details and semantic features of MR-guidance image more effectively by introducing encoder-decoder and deep decoder subnetworks. Moreover, the specific shapes and patterns of the guidance image do not affect the denoised PET image, because the guidance image is input to the network through an attention gate. In a Monte Carlo simulation of [18F]fluoro-2-deoxy-D-glucose (FDG), the proposed method achieved the highest peak signal-to-noise ratio and structural similarity (27.92 ± 0.44 dB/0.886 ± 0.007), as compared with Gaussian filtering (26.68 ± 0.10 dB/0.807 ± 0.004), image guided filtering (27.40 ± 0.11 dB/0.849 ± 0.003), deep image prior (DIP) (24.22 ± 0.43 dB/0.737 ± 0.017), and MR-DIP (27.65 ± 0.42 dB/0.879 ± 0.007). Furthermore, we experimentally visualized the behavior of the optimization process, which is often unknown in unsupervised CNN-based restoration problems. For preclinical (using [18F]FDG and [11C]raclopride) and clinical (using [18F]florbetapir) studies, the proposed method demonstrates state-of-the-art denoising performance while retaining spatial resolution and quantitative accuracy, despite using a common network architecture for various noisy PET images with 1/10th of the full counts. These results suggest that the proposed MR-GDD can reduce PET scan times and PET tracer doses considerably without impacting patients.

deep learning, Deep Image Prior, Attention

1. **Introduction**

Positron emission tomography (PET) is a functional imaging modality that observes the molecular-level activity in tissues caused by radioactive tracers. It offers excellent diagnostic accuracy both for observing normal tissues and for detecting specific diseases such as cancer and neurodegenerative disorders (Phelps, 2012). In response to the increased demand for more accurate dementia diagnosis in recent years, brain-dedicated PET scanners with enhanced sensitivity that are capable of acquiring high-resolution brain images have been developed (Tashima et al., 2019; Watanabe et al., 2017). The acquisition of high-quality diagnostic PET images requires the administration of high dose or a long scan time. However, massive radiation exposure to PET tracers may induce genetic damage and cancerous growths, thereby raising health risk concerns (ICRP, 2017). Therefore, to mitigate the radiation exposure-related risk, it is desirable to administer low-dose PET tracers. Unfortunately, this increases the statistical noise, thus degrading the quality of PET images and potentially affecting the diagnostic accuracy. Thus, improved noise suppression methods for PET images are essential.

Conventionally, Gaussian filtering (GF) is applied as a basic post-denoising method, despite compromising the spatial resolution and, thus, the quantitative accuracy of PET images. To avoid such compromises, various denoising algorithms, such as bilateral filtering (Hoifheinz et al., 2011), non-local means filtering (Arabi and Zaidi, 2020), image-guided filtering (IGF) (He et al., 2013; Hashimoto et al., 2018), and block-matching filtering (Ote et al., 2020) have been developed and applied to PET images. These post-denoising algorithms provide a better denoising performance than GF while retaining spatial resolution and quantitative accuracy.

Aside from the above mentioned conventional filtering algorithms, methods based on deep learning (DL) have been applied in various medical fields (Litjens et al., 2017), and the use of convolutional neural networks (CNNs) has been reported to improve the quality of PET images (Gong et al., 2019b; Häggström et al., 2019; Liu and Qi, 2019; Sanaat et al., 2020; Sphuler et al., 2020; Zhou et al., 2020). However, the general CNN-based denoising methods typically require a pair of large reference datasets comprising high-quality images. This is a major problem in clinical usage owing to the difficulty of preparing huge sets of low-noise PET data without unduly burdening patients. In addition, the huge volume of novel PET tracers being developed, makes it difficult to collect large amounts of training data for each domain. The interpretation of denoised PET images may suffer inherent biases if unknown cases are excluded from the training dataset. Despite these challenges, DL algorithms often outperform conventional denoising algorithms. Therefore, there is a need for technology that can be uniformly adapted to various domains without using high-quality PET data.

In recent years, unsupervised or self-supervised DL approaches such as Noise2Noise and deep image prior (DIP) have demonstrated the potential to overcome these challenges (Lehtinen et al., 2018; Ulyanov et al., 2018). In particular, the DIP algorithm is a powerful noise suppression method that does not require the preparation of a prior training dataset (Hashimoto et al., 2019; Hashimoto et al., 2020; Lin and Huang, 2020). Furthermore, PET reconstruction and denoising methods in which computed tomography (CT) and magnetic resonance (MR) images serve as the prior images input to the DIP framework have been developed (Cui et al., 2019; Gong et al., 2019a). Compared to PET alone, the denoising performance has been improved by using multi-modal data combined with anatomical information. However, while this method shows potential for adapting various PET image denoising approaches, the network architecture does not fully utilize the semantic features or image details of anatomical-guidance images. Furthermore, the process of converting the guidance image to the PET image can result in the shape and pattern of the guidance image remaining in the output PET image, with the extent to which the features of the guidance image affect PET image denoising is yet to be elucidated. Moreover, the process by which the DIP algorithm optimizes the denoising remains unclear. Therefore, these issues must be clarified for future development.

In this study, we propose an unsupervised 3D PET image denoising method that incorporates anatomical information into the DIP architecture via an attention mechanism. The attention gates (Fukui et al., 2019; Schlemper et al., 2019) used in the proposed network help optimize a noisy PET image using multi-scale semantic features extracted from the guidance image. As such, this method can prevent the leakage of guidance image features. The guidance of multi-scale features can lead to an effective regularizer for PET image denoising. The main contributions of this study are as follows:

- We propose a new PET image denoising method guided by anatomical information using an unsupervised DL method.
- We demonstrate that the proposed network has the flexibility to handle the different PET tracer domains used to show the distribution of various tissues in the brains of human and non-human primates.
- The behavior of the optimization process is visualized experimentally, thereby providing useful insights that were unresolved in unsupervised CNN-based restoration problems.

## 2. Related work

The PET image restoration process is further complicated by the limited availability of information that can be extracted from noisy PET data. Another approach adopted for PET image denoising is to use anatomical prior extracted from the CT or MR images of the patient for regularization. Because multi-modal images have become easier to obtain owing to the increasing availability of PET/CT and PET/MR scanners, various hybrid methods using anatomical priors have been developed to facilitate PET image denoising.

## 2.1. Classical approach

Conventionally, hybrid denoising methods using anatomical priors have been adopted for PET image reconstruction and post-filtering. For example, maximum a posteriori image reconstruction has been incorporated alongside anatomical priors (Comtat et al., 2002; Vunckx et al., 2012). In addition, Sudarshan et al. (2020) proposed joint PET and MR image reconstruction using a patch-based joint-dictionary prior. Bland et al. (2018) introduced MR-derived kernels to the kernel expectation maximization reconstruction. Although advanced image reconstruction algorithms can provide better denoising performance, they require significant computational resources and it is often difficult to set optimal parameters. Therefore, anatomically-guided post-denoising is often performed separately from the image reconstruction process. Chan et al. (2014) proposed incorporating CT information and applying median non-local mean filtering to achieve PET denoising. Alternatively, Yan et al. (2015) proposed MR-guided PET filtering by adapting a local linear model. In addition, the authors performed partial volume correction without MR image parcellation by incorporating partial volume effects into the model.

## 2.2. Supervised DL approach

The supervised DL approach has recently demonstrated state-of-the-art performance in PET image denoising. When large amounts of training and label PET data pairs are available, the general CNN can be trained according to the following operation:

$$\theta^* = \operatorname*{argmin}_{\theta} \frac{1}{N_t} \sum_{i \in D_t} \left\| x_{\text{ref}}^i - f_\theta(x_0^i) \right\|, \qquad (1)$$

where $\|\cdot\|$ is the L2 norm, $f$ represents the CNN, $\theta$ denotes the trainable parameters contained in the weights and biases, $D_t$ is a mini-batch sample of size $N_t$, $x_0^i$ is the *i*-th element in the training data (noisy PET images), and $x_{\text{ref}}^i$ is the *i*-th element in the label data (clean PET images). Regarding supervised PET image denoising using anatomical information, in separate studies, Liu and Qi et al. (2019), Schramm et al. (2021), and Chen et al. (2019) trained CNNs to map multi-modal images, including noisy PET and MR images, to obtain clean PET images. In accordance with these methods, the mapping function, $f_\theta(x_0^i)$, in Eq. 1 is represented by $f_\theta(x_0^i, g^i)$ using anatomical-guidance images, $g$. The supervised DL approach requires a vast number of low-dose (or short-time scan), high-dose (or long-time scan), and guidance image pairs.

## 2.3. Unsupervised DL approach

Unsupervised DL approaches such as DIP do not require label data for PET image denoising. The DIP training process is optimized as follows (Ulyanov et al., 2018):

$$\theta^* = \underset{\theta}{\mathrm{argmin}}\|x_0 - f_\theta(z)\|, \quad x^* = f_{\theta^*}(z), \tag{2}$$

where $x_0$ is a noisy image, $x^*$ is the final denoised image output, and the network input $z$ is random noise. The DIP algorithm uses a CNN to map a degraded image, $x_0$, and obtains the optimal denoised image by regularization of the architecture via moderate iteration. This is based purely on the prior information included in the CNN structure. Hashimoto et al. (2019; 2020) proposed dynamic PET image denoising by directly inputting static PET images into 3D and 4D DIP as prior information. In contrast, Cui et al., (2019) and Gong et al. (2019a) both proposed PET denoising methods that rely on inputting anatomical-guidance images, $g$ (e.g., CT and MR images), instead of the DIP input, $z$. In these methods, the mapping function, $f_\theta(z)$, in Eq. 2 is represented by $f_\theta(g)$.

The unsupervised DL approach can bridge the technical gap between the classical and supervised DL approaches for PET image denoising based on anatomical information. Nevertheless, previous methods often fail to clarify how the semantic features of the guidance image affect PET image denoising.

## 3. Methodology

### 3.1. MR-guided deep decoder

To explicitly utilize the semantic features of anatomical-guidance image for PET image denoising, we propose a method for unsupervised 3D PET image denoising based on anatomical information that uses an MR-guided deep decoder (MR-GDD), which is inspired by Uezato et al. (2020). Figure 1 shows the network structure of the proposed MR-GDD, which comprises two subnetworks: an encoder-decoder subnetwork with a skip connection and a deep decoder subnetwork. The two subnetworks are connected by an upsampling refinement unit (URU) and a feature refinement unit (FRU), which incorporate an attention gate to weight the multi-scale features extracted from the MR image to the deep decoder subnetwork. This combined network performs end-to-end learning from scratch.

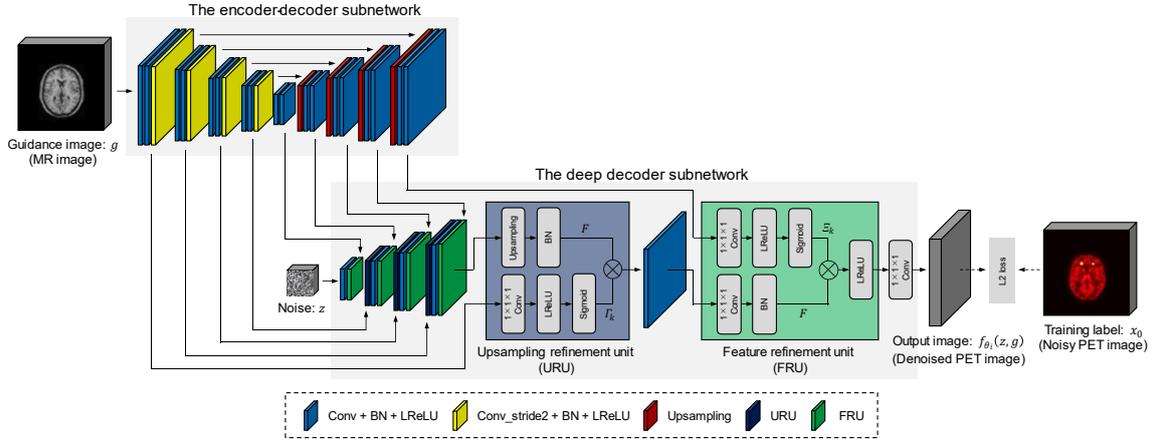

Fig. 1. Overview of the proposed MR-GDD used for unsupervised 3D PET image denoising. This architecture consists of two subnetworks connected by attention gates: an encoder-decoder subnetwork with a skip connection (top) and a deep decoder subnetwork (bottom). This combined architecture is optimized using end-to-end learning. The attention gates in the upsampling refinement unit (URU) and the feature refinement unit (FRU) guide the optimization of noisy PET image using the multi-scale semantic features extracted from the MR image. This architecture can prevent the semantic features of the MR image from leaking.

### 3.1.1. Encoder-decoder subnetwork

The encoder-decoder subnetwork is designed to extract low-scale to high-scale hierarchical semantic features from the MR image. It is based on the 3D U-Net architecture (Çiçek et al., 2016) and consists of encoding and decoding paths. In the encoding path, the combination of a 3 × 3 × 3 3D convolution layer with batch normalization (BN) and a leaky rectified linear unit (LReLU) is repeated twice, before being constructed by a 3 × 3 × 3 3D convolution layer with two strides for downsampling, followed by the BN and LReLU. At each downsampling step, the size of the feature maps is halved. In the decoding path, the outputs of the upsampling layer and the skip connection supplied from the encoding path are added, before the combination of the 3 × 3 × 3 3D convolution layer with the BN and LReLU is repeated twice. At each upsampling step, the size of the feature maps is doubled.

### 3.1.2. Deep decoder subnetwork

The deep decoder subnetwork reconstructs the denoised PET image from the network input filled with uniform noise. Each step in the deep decoder subnetwork is upsampled first by the URU, then by a 3 × 3 × 3 3D convolution layer, and finally by the BN, LReLU, and FRU. Owing to the presence the attention gates, the URU and FRU can generate conditional weights using the MR image features via a 1 × 1 × 1 3D convolution layer, LReLU, and sigmoid function, and then weights the features obtained from the pre-layer in the deep decoder. The URU promotes the spatial locality of MR image features, whereas the FRU promotes similar semantic alignment. Finally, a 1 × 1 × 1 3D convolution layer outputs a denoised PET image.

### 3.2. Loss function and optimization

To denoise the PET image, the training process of the proposed MR-GDD, which utilizes unsupervised learning that does not require PET reference data, is optimized as follows:

$$\theta^* = \underset{\theta}{\mathrm{argmin}} \|x_0 - f_\theta(z,g)\|, \ x^* = f_{\theta^*}(z,g), \tag{3}$$

where $\|\cdot\|$ is the L2 norm, $f$ represents the proposed MR-GDD network, the training label $x_0$ represents the noisy PET image, and the network inputs $z$ and $g$ are the random noise and the MR-guidance image, respectively. The input noise was generated by adding a fixed uniform random noise and Gaussian noise with different seed for each epoch. The attention gates used in the proposed architecture help optimize the noisy PET image by using the multi-scale semantic features extracted from a guidance image $g$. If the *k*-th scale feature of the encoder path in the encoder-decoder subnetwork is defined as $\Gamma_k$ and the *k*-th scale features of the decoder path are defined as $\Xi_k$, the influence on the mapping function, $f$, can be expressed as follows:

$$f = f_\theta(z|\Gamma_1, \cdots \Gamma_k, \Xi_1, \cdots \Xi_k). \tag{4}$$

In this method, the limited memory BFGS (L-BFGS) algorithm (Zhu et al., 1997), which is a quasi-Newtonian method, is introduced to solve the nonlinear least squares problem described by Eq. 3. By considering the approximate Hessian matrix based on the second-order gradient, the L2 norm converges more stably and quickly than first-order gradient descent algorithms, such as the stochastic gradient descent algorithm or Adam. In addition, because a small amount of data is generated as a result of using the unsupervised architecture, which does not require a prior training dataset, the MR-GDD can reduce the computational complexity and load required to performed denoising. We used the L-BFGS algorithm at a learning rate of 0.01 without line search to minimize the processing time. The proposed architecture was processed using PyTorch 1.6.0 on Ubuntu 16.04 with acceleration by a graphics processing unit (NVIDIA Quadro RTX8000 with 48 GB memory).

### 4. Experimental setup

A simulation study using [$^{18}$F]fluoro-2-deoxy-D-glucose (FDG), a preclinical study using [$^{18}$F]FDG and [$^{11}$C]raclopride, and a clinical study using [$^{18}$F]florbetapir (Wong et al., 2010) were performed to verify the effectiveness of the proposed PET image denoising method. In addition, the denoising performance of the proposed MR-GDD was compared against four other unsupervised algorithms under the same conditions as the proposed MR-GDD;

- **Gaussian filtering (GF).** GF is a basic post-denoising method for suppressing noise in PET images. We used a 3D Gaussian kernel.
- **Image guided filtering (IGF).** IGF performs PET image denoising by adapting a local linear model using a guidance image (Hashimoto et al., 2018; He et al., 2013). We used the MR image as the guidance image.
- **Deep image prior (DIP).** The general DIP algorithm uses random noise as the network input (Ulyanov et al., 2018). We used the encoder-decoder network in the proposed MR-GDD as the DIP architecture.
- **MR-DIP.** MR-DIP uses an MR prior as a direct input in the DIP architecture (Cui et al., 2019; Gong et al., 2019a). We used the same architecture as in the DIP algorithm.

As a pre-processing step for all the data, voxel intensity normalization was performed on each of the MR and PET images. The MR image intensity was normalized in the [0, 1] range using a min-max normalization technique. Furthermore, the 99.95th percentile was defined for PET image intensity, which was also normalized in the [0, 1] range using a min-max normalization technique.

## 4.1. Simulation study

We performed a Monte Carlo simulation using the 3D brain phantom from BrainWeb (Aubert-Broche et al., 2006). The Monte Carlo simulation modeled the geometry of a brain-dedicated PET scanner (HITS-655000 (Watanabe et al., 2017), Hamamatsu Photonics K.K., Japan). The scanner consists of 32 detector blocks with cerium doped lutetium yttrium orthosilicate crystals per ring, with 5 of these blocks aligned to the scanner axis. To ensure a mismatch between the MR and PET images, two hot spheres (with radii of 10 and 12 mm, respectively) were inserted exclusively into the PET image to present tumor regions. Then, we simulated a static [$^{18}$F]FDG scan equivalent to 150M counts, including attenuation and scattering effects. An attenuation map was created from the segmented MR image, and the attenuation effects of water and bone were considered. The gray matter:white matter:cerebrospinal fluid:left tumor:right tumor activity ratio was set to 1:0.25:0.05:2:2 based on the [$^{18}$F]FDG contrast. We also simulated low-activity tumor data (left tumor:right tumor activity ratio was set to 1.5:1.2) for some low-grade tumors or pathology linked to neurodegenerative disease. To generate the reference PET image, the simulated list-mode data were reconstructed using a 3D list-mode dynamic row-action maximum-likelihood algorithm (DRAMA) (Tanaka and Kudo, 2010) with two iterations and 40 subsets. The reconstructed image measured 128 × 128 × 83 voxels (2.6 × 2.6 × 2.4 mm/voxel). The reference image was reconstructed using all the list-mode data. The noisy PET image was obtained by periodically downsampling to 1/30th of the reference list-mode data (see Supplementary Material 0.1) and reconstructing the low-count image (5M counts). The corresponding T1-weighted MR image was used as the guidance image.

### 4.2. Preclinical study

A preclinical study was conducted using monkeys and was approved by the Animal Ethics Committees of the Central Research Laboratory, Hamamatsu Photonics K.K. PET scans using [$^{18}$F]FDG and [$^{11}$C]raclopride imaged the brains of conscious rhesus monkeys, whose bodies and heads were fixed using an animal PET scanner (SHR-38000 (Hamamatsu), Hamamatsu Photonics K.K., Japan). After a 30 min transmission scan using a $^{68}$Ge-$^{68}$Ga source, doses of [$^{18}$F]FDG (113 MBq) and [$^{11}$C]raclopride (282 MBq) were injected into each monkey, before dynamic PET scans were performed for 120 and 90 min, respectively. We performed image reconstruction using 3D DRAMA with two iterations and 60 subsets that incorporated attenuation correction via transmission scan data. The reconstructed images measured 256 × 256 × 103 voxels (0.65 × 0.65 × 1.0167 mm/voxel) and were cropped to 192 × 192 × 64 voxels to reduce the demand on the GPU memory. The reference images were reconstructed using all the list-mode data for [$^{18}$F]FDG and the list mode data for the 60 min period between 30 and 90 min of the scan such that the contrast created by the distribution of [$^{11}$C]raclopride in the striatum to be observed clearly. Each noisy PET image was obtained by periodically downsampling to 1/10th of the reference list-mode data. The corresponding T1-weighted MR images were taken on another day and registered manually by two radiological technologists.

### 4.3. Clinical study

An amyloid scan using [$^{18}$F]florbetapir was conducted on the human brain of a cognitive normal subject using a Biograph mMR scanner (Siemens Healthineers, Germany) as part of the "Insight 46" sub-study of the MRC National Survey of Health and Development (Lane et al., 2017; Markiewicz et al., 2018a). The PET data were acquired dynamically for 60 min using a [$^{18}$F]florbetapir dose of 409 MBq. During PET image reconstruction, attenuation correction was performed via a pseudo-CT image, which was synthesized using T1- and T2-weighted MR images that were acquired simultaneously, and enabled an μ-map to be was calculated. Image reconstruction was performed using an ordered subset expectation maximization (OS-EM) algorithm involving four iterations and 14 subsets using the NiftyPET package (Markiewicz et al., 2018b). The reconstructed images comprised 334 × 334 × 127 voxels (2 × 2 × 2 mm/voxel), which were cropped to 128 × 128 × 83 voxels to reduce the demand on the GPU memory. The reference image was reconstructed using the list-mode data for the period from 30 min to 60 min to observe the contrast between gray and white matter clearly. The noisy PET image was obtained by periodically downsampling to 1/10th of the reference list-mode data.

**4.4. Evaluation metrics**

To quantitatively evaluate the performance of different denoising methods, for the simulation study, we calculated the peak signal-to-noise ratio (PSNR) and structural similarity (SSIM) of the target image, $x$, and the reference image, $x^{\text{ref}}$, which are defined as

$$\text{PSNR} = 10 \log_{10} \left( \frac{\max(x^{\text{ref}})^2}{\frac{1}{N_R} \sum_{j \in R} (x_j - x_j^{\text{ref}})^2} \right) \quad (5)$$

and

$$\text{SSIM} = \frac{1}{N_R} \sum_{j \in R} \frac{\left(2\mu_{jx}\mu_{jx^{\text{ref}}} + c_1\right)\left(2\sigma_{jxx^{\text{ref}}} + c_2\right)}{\left(\mu_{jx}^2 + \mu_{jx^{\text{ref}}}^2 + c_1\right)\left(\sigma_{jx}^2 + \sigma_{jx^{\text{ref}}}^2 + c_2\right)}, \quad (6)$$

respectively. Here, $R$ and $N_R$ represent the brain region in the PET image and the number of voxels, respectively, $\mu$ and $\sigma$ are the mean and standard deviation of the square window corresponding to the $j$-th voxel, respectively, and $\sigma_{jxx^{\text{ref}}}$ is the covariance between $x$ and $x^{\text{ref}}$. Furthermore, $c_1 = (0.01L)^2$ and $c_2 = (0.03L)^2$ where $L$ represents the dynamic range of the reference. The contrast-to-noise ratio (CNR) between gray matter and white matter was calculated as

$$\text{CNR} = \frac{|\overline{S_g} - \overline{S_w}|}{\sqrt{\sigma_g^2 + \sigma_w^2}}, \quad (7)$$

where $\overline{S_g}$, $\overline{S_w}$ and $\sigma_g$, $\sigma_w$ are the mean activity and standard deviation corresponding to the regions of interest (ROIs) in the gray and white matter, respectively. The ROIs were constructed in the gray matter regions on the left and right brains of different slices, whereas the white matter regions were selected within the centrum semiovale (Paxinos et al., 1999). In the [$^{11}$C]raclopride calculation, the ROIs were set on the putamen instead of the gray matter. A Wilcoxon signed-rank test was performed on the PSNR, SSIM, and CNR to compare the performance of different denoising methods. As the noisy PET data were generated by downsampling the reference list-mode data, we generated multiple independent noisy samples (10 samples) to examine the uncertainty and test the statistical significance.

We also evaluated the tradeoffs between noise and quantitative information of the denoised images when FWHM (GF), $\varepsilon$ (IGF), and epochs (CNNs) are changed. For the simulation study, we calculated the tradeoff between the mean squared bias and the variance in the ROI:

$$\text{Bias}^2 = \frac{\sum_{j \in R}(x_j - x_j^{\text{ref}})^2}{\sum_{j \in R}(x_j^{\text{ref}})^2}, \tag{8}$$

$$\text{Variance} = \frac{\sum_{j \in R}(x_j - \bar{x})^2}{\sum_{j \in R}(x_j^{\text{ref}})^2}, \tag{9}$$

where R denotes the tumor regions, $\bar{x}$ is the average pixel value inside the ROI. For preclinical study, we calculated the tradeoff between the mean uptake (putamen and caudate and striatum) and the standard deviation (white matter). Each ROI used for the evaluation was set manually to include any partial volume voxels on the MR image and was calculated via superimposition on the co-registered PET image.

Thus far, the optimization process for the DIP algorithm has only been reported as a conceptual diagram (Hashimoto et al., 2020; Ulyanov et al., 2018), with several aspects such as the actual behavior remaining unclarified. To elucidate the DIP optimization process, we visualized it by performing nonlinear dimensionally reduction using the locally linear embedding (LLE) algorithm (Saul and Roweis, 2003), which implements manifold learning. We projected the PET data used for optimization under different training conditions onto a three-dimensional manifold, considering a neighborhood of 15 for each point.

## 5. Results
### 5.1. Simulation study

Figure 2 shows simulations of three orthogonal slices processed by different denoising algorithms. In addition, Table 1 shows the results of a statistical analysis of the PSNR, SSIM, and CNR for different denoising methods. The proposed MR-GDD yielded higher values for PSNR, SSIM, and CNR. Furthermore, a statistical significance ($p$ value < 0.05) was noted between MR-DIP and MR-GDD for all the quantitative evaluation values (see Supplementary Material 0.2.1). Compared with the alternative denoising algorithms, by visual inspection, the proposed MR-GDD achieved a clearer restoration of the smallest inserted tumor. The corresponding line profiles passing through both tumors are compared in Fig. 3. The proposed MR-GDD retrieved the highest contrast in the area surrounding the smallest inserted tumor. Figure 4 illustrated the shows noise-bias tradeoff when the processing conditions used for the different denoising algorithms are changed. The proposed MR-GDD exhibits a lower mean squared bias and variance than the other denoising algorithms.

Figure 5 shows attention maps corresponding to the different channels used in the URU extracted from the first scale encoder path and the FRU extracted from the fifth (final) scale decoder path. The conditional weights showing the image details and semantic features were generated as attention maps.

Figure 6 shows the optimization processes of the DIP, MR-DIP, and MR-GDD algorithms projected onto a three-dimensional manifold by the LLE algorithm. The trajectory of the proposed MR-GDD optimization process corresponded most closely to the reference point.

Figure 7 shows the processed images and line profiles of noisy PET images created by changing the rate of thinning out the counts when using low-activity simulated tumors like low grade tumor. The proposed MR-GDD achieved a clearer restoration of the smallest low-activity tumor.

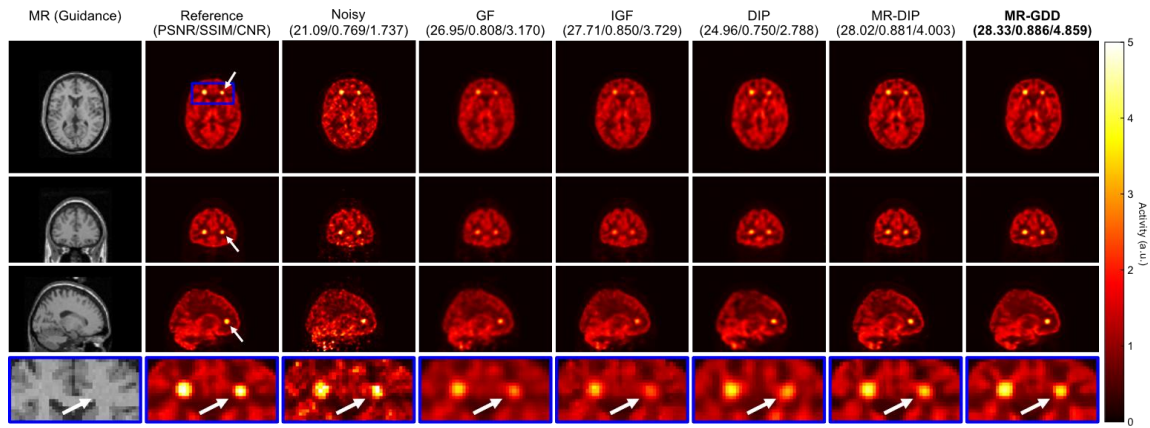

Fig. 2. Three orthogonal slices processed by applying different denoising algorithms to static brain-PET images simulated using Monte Carlo modeling. From left to right, the columns show sample images of the T1-weighted MR guidance, reference (150M counts), noisy (5M counts), and the denoised PET images processed using GF, IGF, DIP, MR-DIP, and the proposed MR-GDD algorithms. The white arrows in the reference and magnified images indicate the smallest inserted tumors. The corresponding PSNR, SSIM, and CNR values are shown in brackets above the columns showing processes images. The bottom row shows the magnified images of the area marked by a blue rectangle. The proposed MR-GDD achieved a clearer restoration of the smallest inserted tumor (indicated by white arrows in magnified images).

Table 1

Results of quantitative evaluation and statistical analysis used in the simulation study. Bold values indicate the best method for each quantitative evaluation.

| Method | PSNR [dB] | | SSIM | | CNR | |
|---|---|---|---|---|---|---|
| | mean ± std | $p$ value | mean ± std | $p$ value | mean ± std | $p$ value |
| Noisy | 20.73±0.14 | < 0.001 | 0.792±0.041 | < 0.001 | 1.798±0.202 | < 0.001 |
| GF | 26.68±0.10 | < 0.001 | 0.807±0.004 | < 0.001 | 3.246±0.299 | 0.002 |
| IGF | 27.40±0.11 | 0.003 | 0.849±0.003 | < 0.001 | 3.886±0.480 | 0.138 |
| DIP | 24.22±0.43 | < 0.001 | 0.737±0.017 | < 0.001 | 2.502±0.565 | < 0.001 |
| MR-DIP | 27.65±0.42 | 0.003 | 0.879±0.007 | < 0.001 | 3.650±0.672 | 0.041 |
| MR-GDD | **27.92±0.44** | - | **0.886±0.007** | - | **4.153±0.518** | - |

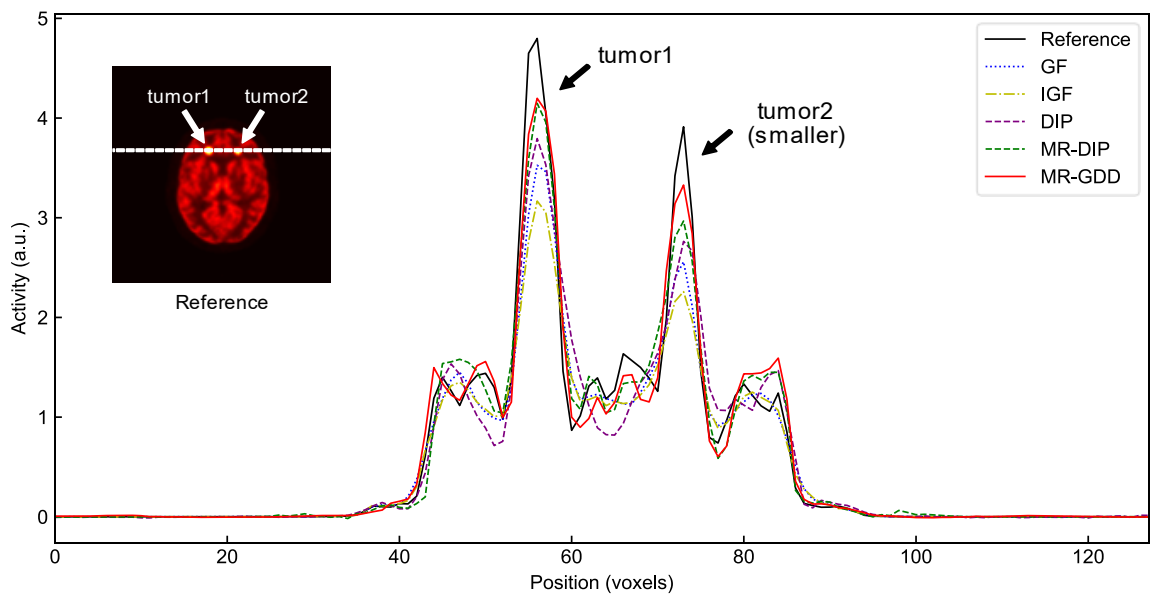

Fig. 3. Comparison of the line profiles passing through both inserted tumors in the denoised images processed by different denoising algorithms. The proposed MR-GDD method retrieved the highest contrast in the area surrounding the smaller inserted tumor (tumor2).

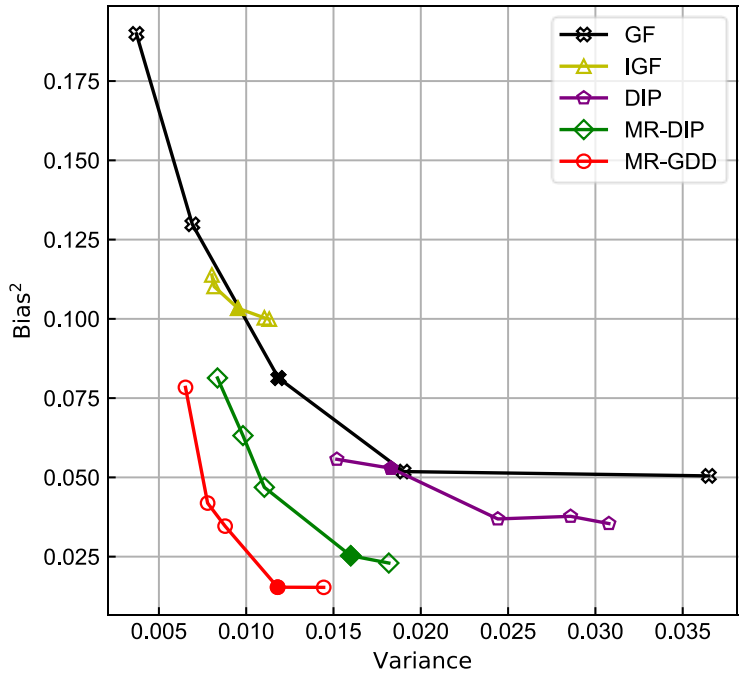

Fig. 4. Mean squared bias versus variance curves for different conditions in the tumor ROIs. From the upper left to lower right, the markers correspond to 1.5, 1.25, 1.0, 0.75, and 0.5 mm FWHM (GF); 1e-02, 1e-03, 1e-04, 1e-05, and 1e-06 $\varepsilon$ (IGF); 73, 77, 79, 82, and 90 epochs (DIP); 15, 16, 17, 21, and 22 epochs (MR-DIP); and 20, 22, 23, 28, and 29 epochs (MR-GDD). The filled markers denote the conditions corresponding to the denoised PET images in Fig. 2.

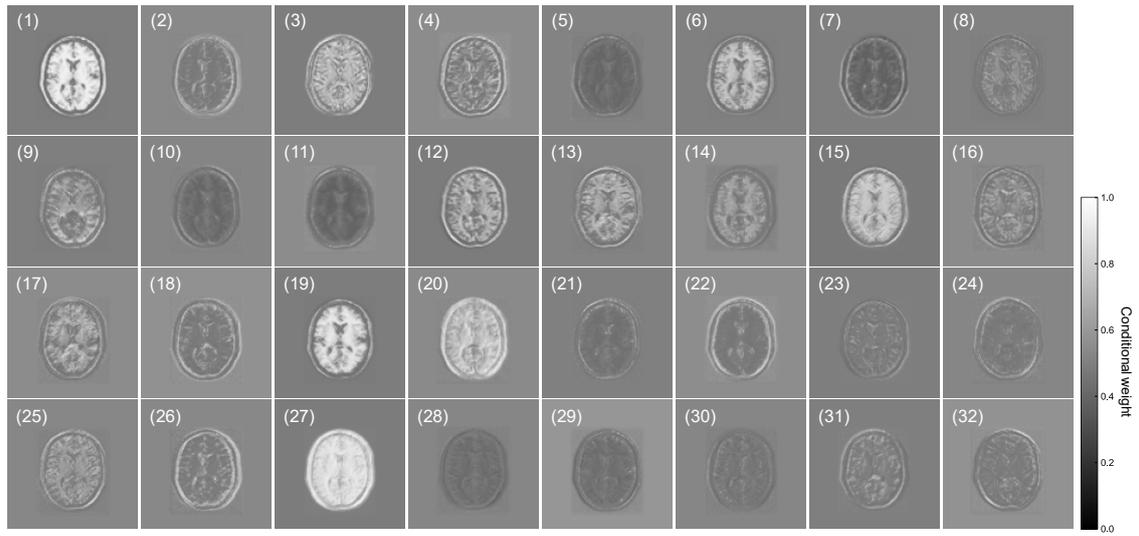

(a) First scale encoder path: $\Gamma_1$

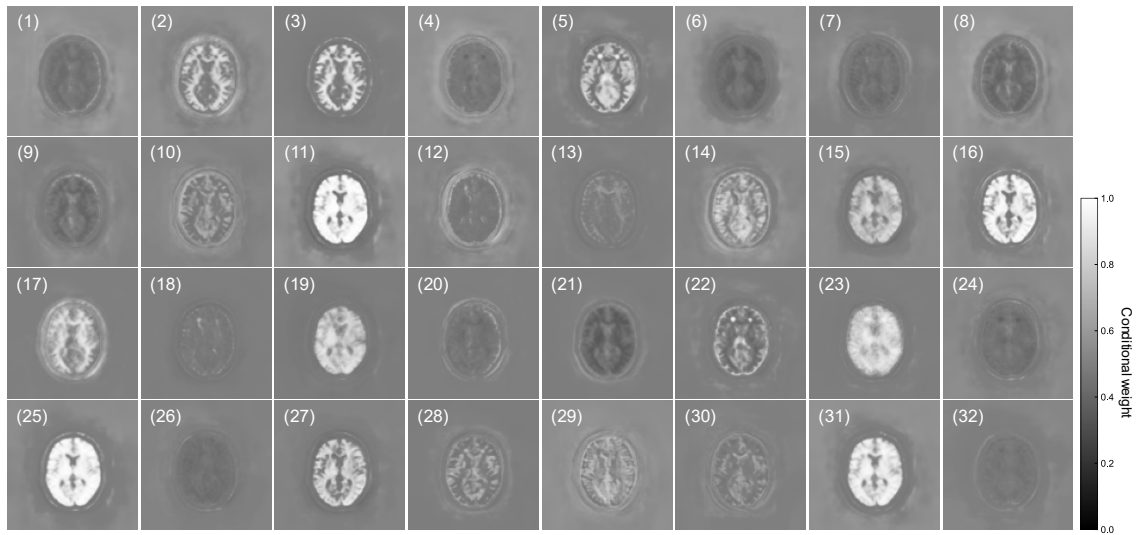

(b) Fifth scale decoder path: $\Xi_5$

Fig. 5. Attention maps corresponding to the different channels used in (a) the URU extracted from the first scale encoder path, $\Gamma_1$, and (b) the FRU extracted from the fifth scale decoder path, $\Xi_5$, with each layer comprising 32 channels. The index in the upper left of each image denotes the channel number. The attention maps showing the spatial details of the MR image were generated in the URU (e.g., (a6)), and the various extracted semantic features indicate the presence of gray matter (e.g., (b10)), white matter (e.g., (b3)), and tumors (e.g., (b22)).

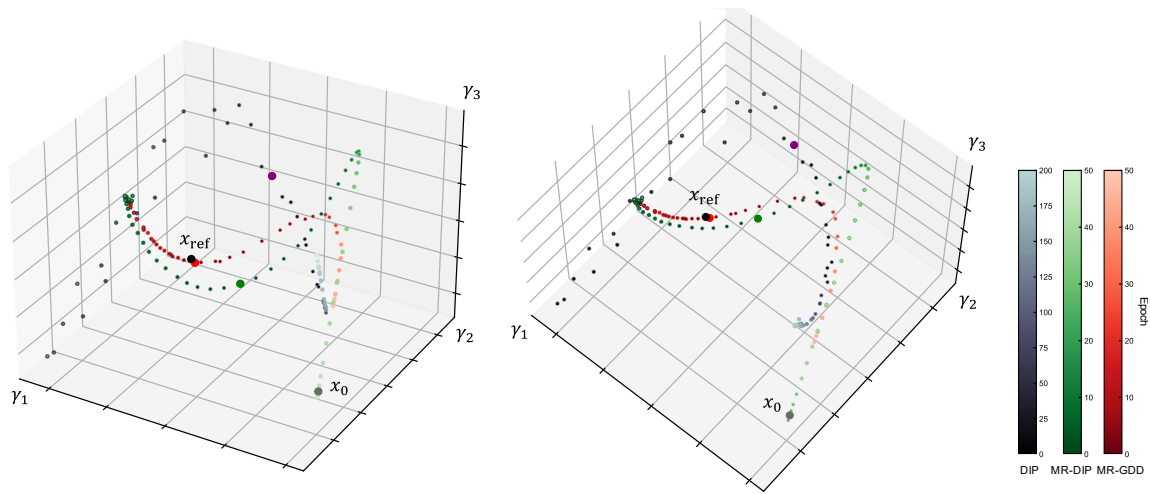

Fig. 6. Visualization of the optimization processes of the DIP (gray), MR-DIP (green), and MR-GDD (red) algorithms projected onto the three-dimensional manifold. The points in the scatter plot show the dimension-reduced output image at each epoch. When these architectures are sufficiently optimized, they converge to a solution (noisy PET data: $x_0$), which is notably distinct from $x_{\text{ref}}$. The larger markers correspond to the denoised PET images in Fig. 2. Each figure shows a projection from a different view angle. Each coordinate position is shown in the Supplementary Material (csv file).

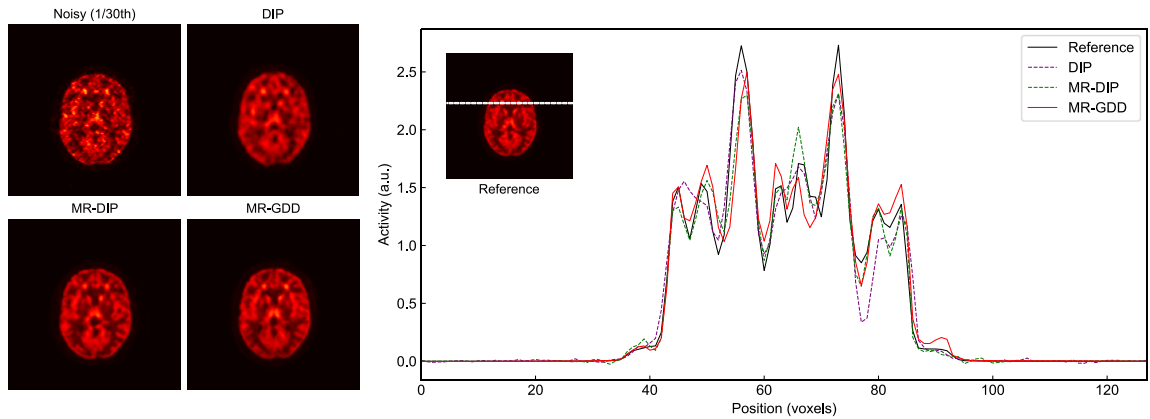
(a) Denoised images and line profiles (1/30th of counts)

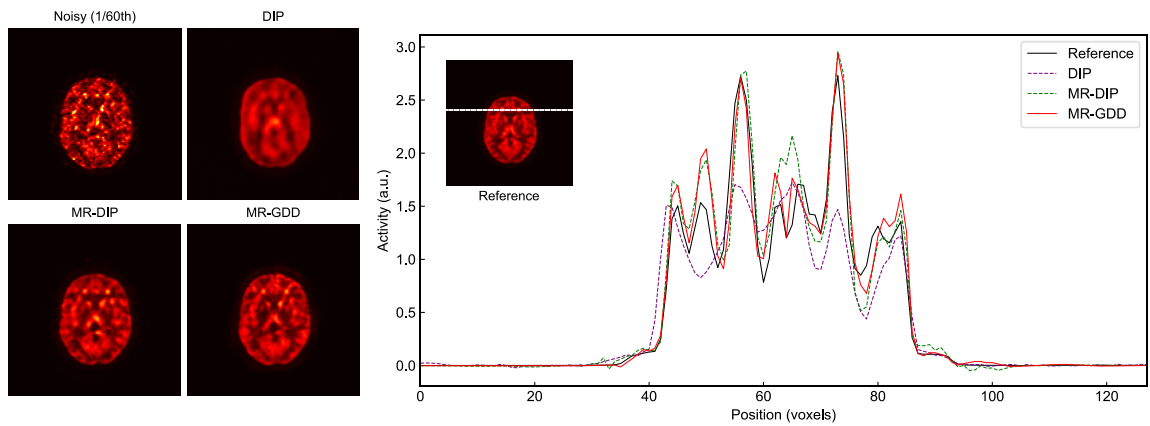
(b) Denoised images and line profiles (1/60th of counts)

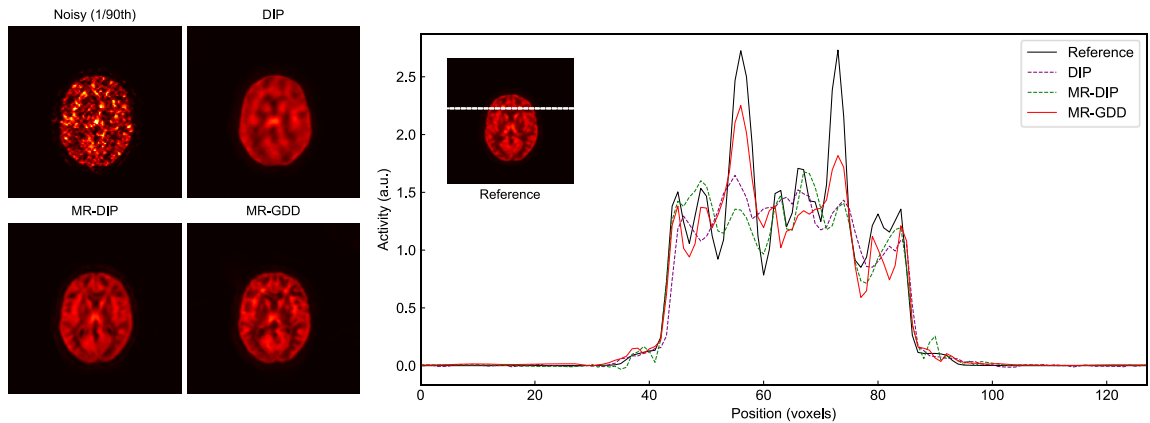
(c) Denoised images and line profiles (1/90th of counts)

Fig. 7. Axial images and line profiles processed by applying different noisy PET images with low-activity tumors, created by changing the rate of thinning of the data counts: (a) 1/30th (5M counts), (b) 1/60th (2.5M counts), and (c) 1/90th (1.7M counts) of counts.

### 5.2. Preclinical study

Figures 8 shows three orthogonal slices processed by different denoising algorithms for

[$^{18}$F]FDG and [$^{11}$C]raclopride. The results of the statistical analysis on [$^{18}$F]FDG show that the mean ± std CNR for the noisy, GF, IGF, DIP, MR-DIP, and the proposed MR-GDD were 2.303 ± 0.177, 3.540 ± 0.242, 3.614 ± 0.263, 2.822 ± 0.317, 3.877 ± 0.151, and 4.196 ± 0.338, respectively. The CNR of the proposed MR-GDD was significantly higher than those of the noisy ($p < 0.001$), GF ($p = 0.002$), IGF ($p = 0.003$), DIP ($p < 0.001$), and MR-DIP ($p = 0.024$) methods. In the case of [$^{11}$C]raclopride, the mean ± std CNR for the noisy, GF, IGF, DIP, MR-DIP, and the proposed MR-GDD were 1.406 ± 0.192, 3.292 ± 0.233, 3.158 ± 0.320, 2.225 ± 0.443, 3.394 ± 0.378, and 3.683 ± 0.401, respectively. The CNR of the proposed MR-GDD was significantly higher than those of the noisy ($p < 0.001$), GF ($p = 0.005$), IGF ($p = 0.002$), DIP ($p = 0.002$), and MR-DIP ($p = 0.024$) methods. The IGF, MR-DIP, and MR-GDD algorithms, which all use MR images outperform the GF and DIP algorithms, with the denoised PET image obtained using the proposed MR-GDD exhibiting the most detailed restoration. Figure 9 shows the noise-uptake tradeoff when the processing conditions in different denoising algorithms are changed for the [$^{18}$F]FDG and [$^{11}$C]raclopride. The proposed MR-GDD method resulted in a higher mean putamen, caudate, and striatum uptake and a lower background standard deviation relative to the other denoising algorithms for both tracers. Therefore, it is suitable for processing images involving both PET tracers; [$^{18}$F]FDG spreads widely and facilitates the extraction of anatomical information, while [$^{11}$C]raclopride accumulates in localized areas of the brain such as the striatum consist of the putamen and caudate, making it difficult to obtain anatomical information, and has a different shape from the guidance image.

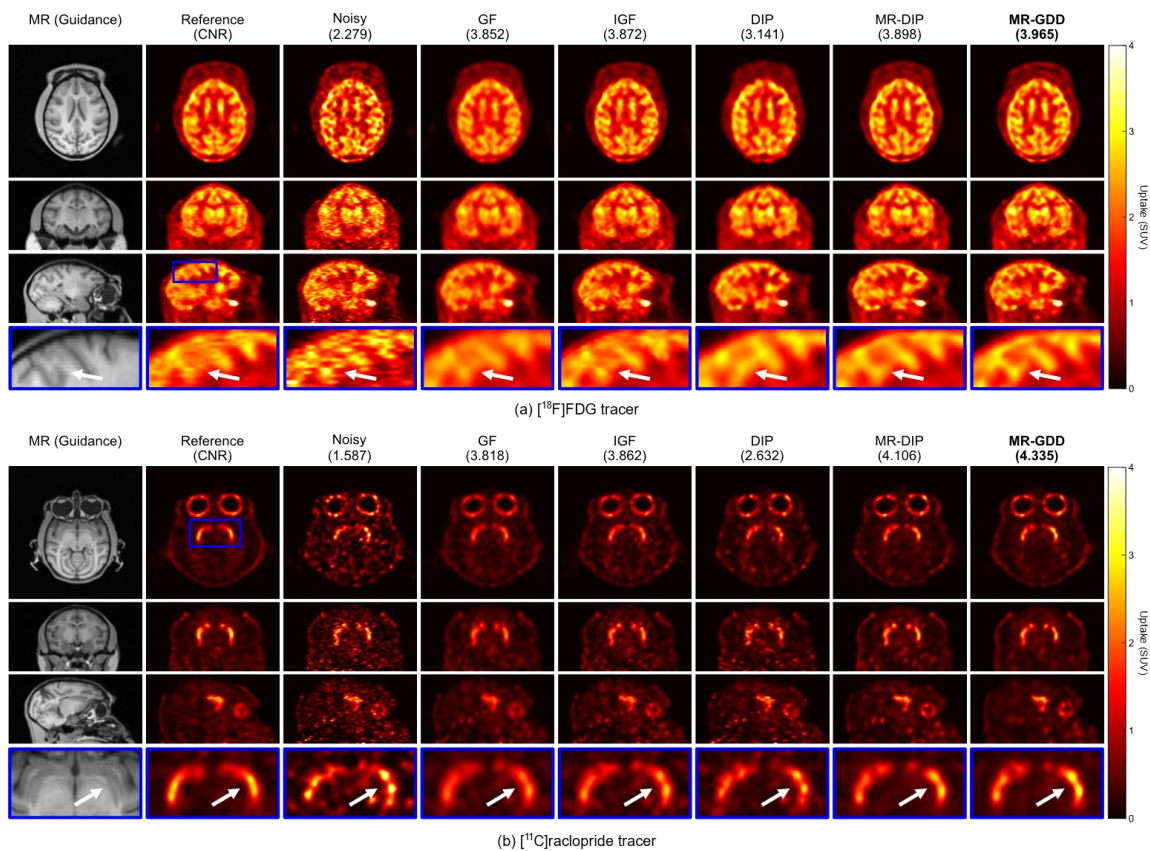

Fig. 8. Three orthogonal slices from PET images of a monkey brain injected with (a) [$^{18}$F]FDG and (b) [$^{11}$C]racloprice that have been processed using different denoising algorithms. From left to right, the sample images represent the T1-weighted MR guidance, reference ((a) accumulated for 120 min, (b) accumulated for 30-90 min), noisy (1/10th of counts), and denoised PET images corresponding to the GF, IGF, DIP, MR-DIP, and proposed MR-GDD algorithms. The corresponding CNR values are shown in brackets above the columns depicting processes images. The bottom row shows the magnified images of the area marked by a blue rectangle. The proposed MR-GDD achieved a clearer restoration of the (a) gray matter and (b) putamen (indicated by white arrows in magnified images).

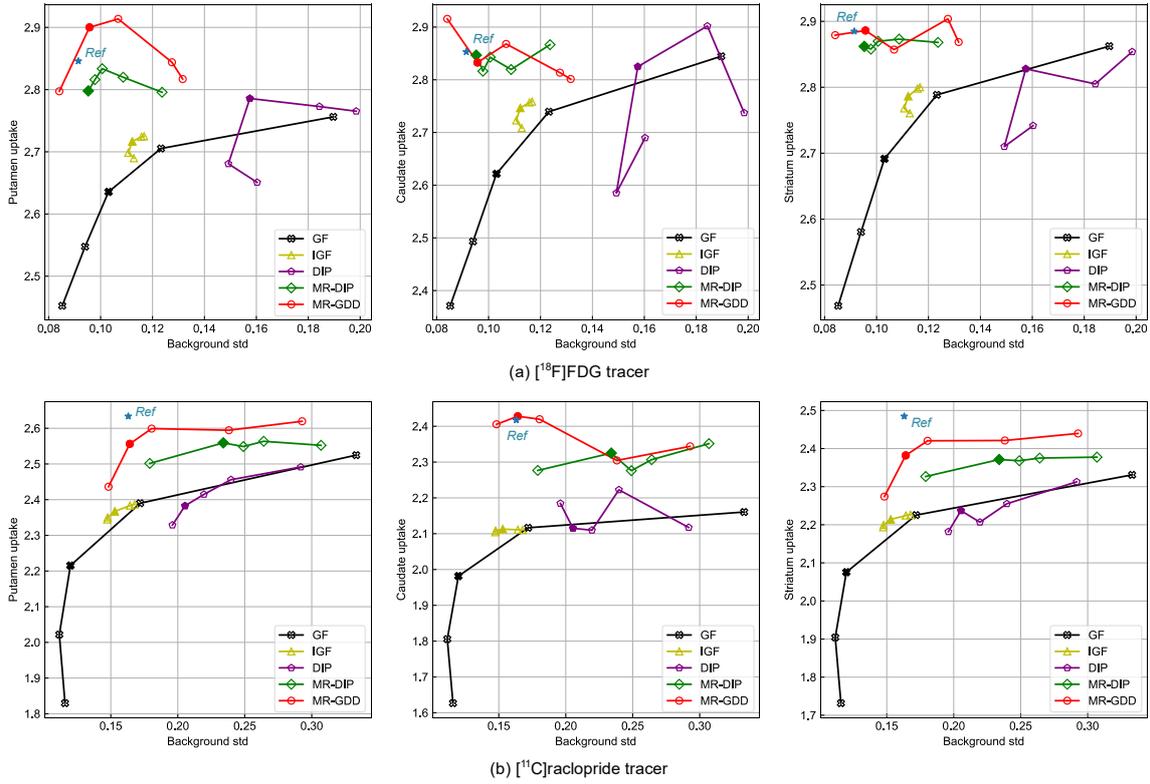

Fig. 9. Curves showing the putamen, caudate, and striatum uptake versus standard deviation for different conditions. For [18F]FDG (left), from left to right for each algorithm, the markers correspond to 2.5, 2.0, 1.5, 1.0, and 0.5 mm FWHM (GF); 1e-02, 1e-03, 1e-04, 1e-05, and 1e-06 ε (IGF); 80, 88, 90, 93, and 96 epochs (DIP); 19, 21, 22, 26, and 28 epochs (MR-DIP); and 45, 46, 62, 71, and 75 epochs (proposed MR-GDD). For [11C]raclopride (right), from left to right for each algorithm, the markers correspond to 2.5, 2.0, 1.5, 1.0, and 0.5 mm FWHM (GF); 1e-03, 5e-04, 1e-04, 1e-05, and 1e-06 ε (IGF); 65, 85, 92, 115, and 119 epochs (DIP); 10, 18, 19, 20, and 21 epochs (MR-DIP); and 23, 25, 26, 32, and 35 epochs (proposed MR-GDD). The reference and denoised PET images in Fig. 8 are labeled using star-shaped and filled markers, respectively.

### 5.3. Clinical study

Figure 10 shows three orthogonal slices processed by different denoising algorithms for the [18F]florbetapir. The PET image processed using the proposed MR-GDD method exhibits the greatest noise removal while preserving edge accuracy. The results of the statistical analysis show that the mean ± std CNR for the noisy, GF, IGF, DIP, MR-DIP, and the proposed MR-GDD were 0.900 ± 0.103, 3.490 ± 0.361, 3.775 ± 0.491, 3.188 ± 0.689, 4.852 ± 0.497, and 5.439 ± 0.680, respectively. The CNR of the proposed MR-GDD was significantly higher than those of the noisy ($p < 0.001$), GF ($p < 0.001$), IGF ($p < 0.001$), DIP ($p < 0.001$), and MR-DIP ($p = 0.042$) methods. A statistical significance ($p$ value < 0.05) was observed for the different denoising methods. The proposed MR-GDD method yielded the highest CNR among the investigated denoising algorithms, providing a clear contrast between gray matter and white matter.

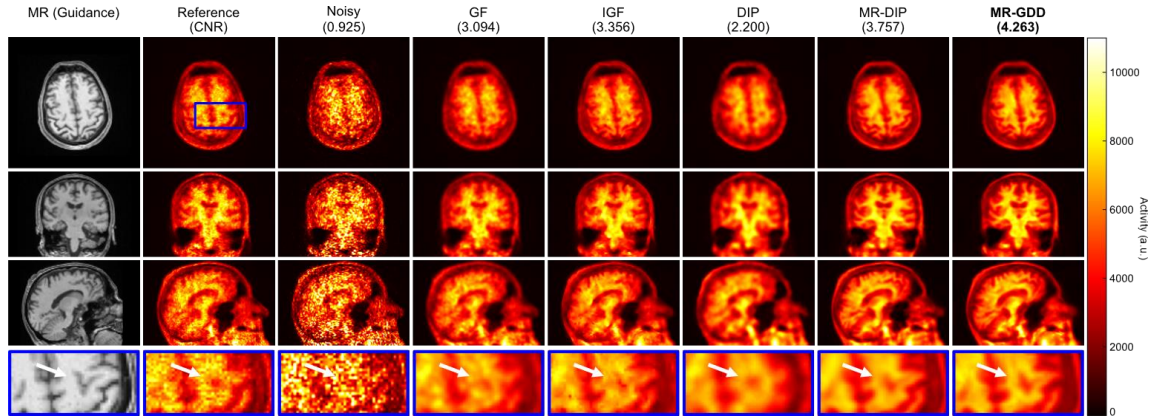

Fig. 10. Three orthogonal slices from PET images of a human brain injected with [$^{18}$F]florbetapir that have been processed using different denoising algorithms. From left to right, the sample images show the T1-weighted MR guidance, reference (accumulated for 30-60 min), noisy (1/10th of counts), and denoised PET images corresponding to the GF, IGF, DIP, MR-DIP, and proposed MR-GDD algorithms. The corresponding CNR values are shown in brackets above the columns displaying processed images. The bottom row shows the magnified images of the area marked by a blue rectangle. The proposed MR-GDD achieved a clearer restoration of the contrast (indicated by white arrows in magnified images).

## 6. Discussion

The proposed MR-GDD method is an unsupervised DL method that provides versatility by using the multi-scale semantic features of MR image for different PET tracers such as [$^{18}$F]FDG, which indicates glucose metabolism; [$^{11}$C]raclopride, which highlights dopamine receptors; [$^{18}$F]florbetapir, which reveals amyloid-β, whereas the general DL method for PET image denoising requires multiple training-label image pairs and must be trained to handle specific tasks.

As shown by Fig. 2 and 4, the IGF, MR-DIP, and MR-GDD algorithms outperformed their GF and DIP counterparts in terms of PSNR, SSIM, CNR, and the bias-variance tradeoff. By using anatomical information as a guidance image, these methods achieve denoising without compromising the spatial resolution and quantitative analysis value of raw PET images. The line profile results (Fig. 3) indicate that the tumor activity calculated using IGF is lower than that calculated using GF. This is attributed to the local filtering that was performed using the MR image as guidance. In contrast, the MR-DIP and MR-GDD algorithms achieved higher activity than the other denoising algorithms. The increased tumor activity yielded by the DIP algorithm is due to the use of non-local information, which distinguishes this approach from the local filtering-based IGF. Indeed, some researchers report the experimental use of the DIP algorithm to perform image denoising for cases involving the self-similarity of the target image (Tachella et al., 2020; Yokoya et al., 2020). The line profile results are consistent with the above-mentioned hypothesis.

As Fig. 2 and 3 demonstrate, the proposed MR-GDD method achieved a clearer restoration of the smallest inserted tumor than the MR-DIP algorithm. In the visualization of the attention map shown in Fig. 5, conditional weights that indicate the spatial details of the MR image were generated on channels (a6), (a12), and (a19) of the URU. Conversely, the FRU generated conditional weights based on a combination of MR and PET image information; various semantic features were extracted, indicating gray matter on channels (b10) and (b28), white matter on channels (b2) and (b3), and tumors on channels (b5) and (b22). It is considered that the features of the MR image were used more appropriately by incorporating the attention mechanism, while excluding the conversion process from the MR image to the PET image. Furthermore, a denoised PET image using only the conditional weights of URU or FRU is shown in Appendix. The performance of the proposed model was superior compared to the network structure composed of only the URU or FRU and that in which the connection of each unit was swapped (see Appendix). The proposed method provides a deep prior that utilizes the spatial details and semantic features of the guidance image more efficiently than in previous studies by weighting the features of the deep decoder network using the URU and FRU (Cui et al., 2019; Gong et al., 2019a; Hashimoto et al., 2019; Hashimoto et al., 2020).

As shown in Fig. 6, we succeeded in visualizing the DIP optimization process experimentally. As reported in previous studies (Hashimoto et al., 2020; Ulyanov et al., 2018), these methods can pass the neighborhood of the reference point along the path to the noisy PET image, while stopping the process early can deliver an improved performance. The DIP, MR-DIP, and MR-GDD methods were successively closer to the reference, and the visualization of the optimization process yielded the same results as the quantitative evaluation in Fig. 4. The initial point of the optimization path differed between the DIP and MR-DIP/MR-GDD algorithms. By using the MR image as a guidance image as opposed to a random noise input, unstable behavior in the initial epoch of the optimization process can be avoided, resulting in an initial point that is closer to reference. Recently, widespread research has attempted to interpret the DIP algorithm, which has many unknown parts. For example, Yokota et al. (2020) reinterpreted the DIP method by using manifold modeling in embedded space (MMSE) and combining a denoising-autoencoder with a multi-way delay-embedding transform. Our results may provide important insights into the development and interpretation of the DIP algorithm.

From the results of the preclinical study shown Fig. 9, the proposed MR-GDD method evidently offers the best noise-uptake tradeoff. The uptake of the MR-GDD passed near the ground truth point. Figure 10 which includes actual clinical data, shows that the CNR of the proposed MR-GDD method was higher than that of the other algorithms, and provided a clear contrast. As the noisy

PET image with 1/10th of counts could be restored to the same quality as the reference data, the proposed method can reduce the PET scan time and the radiation dose of PET tracers significantly without affecting the patients. In addition, because the L-BFGS algorithm can shorten the optimization time (see Supplementary Material 0.2.2), it promises to make a valuable contribution in actual clinical practice.

The limitations of our study and possible future directions of research include the following. Firstly, as PET/CT and PET/MR are currently deployed in most hospitals, multi-modality registration is not required. However, performance deterioration due to patient movement between the PET and CT imaging phases in a PET/CT scan is concern. Therefore, future work should focus on how patient movement affects the performance of the proposed method. Secondly, if the image denoising task is extended to include image restoration, using the MR image may have an effect similar to simultaneous partial volume correction. Many studies have reported that the simultaneous adaptation of denoising and partial volume correction improves performance (Schramm et al., 2021; Xu et al., 2018). We intend to investigate this in future work. Finally, in this study, we comprehensively evaluated PET image denoising using [$^{18}$F]FDG, which indicates glucose metabolism; [$^{11}$C]raclopride, which highlights dopamine receptors; and [$^{18}$F]florbetapir, which reveals amyloid-$\beta$. However, our experiments used only healthy subjects, although tumors were included in the Monte Carlo simulation. From the results of (a) and (b) in the Fig. 7, the proposed MR-GDD had the highest accuracy in terms of the contrast recovery of low-activity tumors. In addition, the proposed method showed some recovery in tumor contrast compared to other methods even when the tumors were almost eliminated by noise, as shown in (c). It is expected to be effective for low-grade tumors or pathology linked to neurodegenerative disease. Future work will evaluate the performance of the proposed MR-GDD algorithm using real data with different and asymmetrical contrasts that represent diseases such as brain atrophy and ischemia.

## 7. Conclusion

In this study, we proposed an unsupervised 3D PET image denoising method that uses anatomical information-guided attention. Our proposed MR-GDD algorithm uses the spatial details and semantic features of an anatomical-guidance image more efficiently. As the guidance image is input to the network via the attention gate, the specific shapes and patterns of the guidance image do not affect the denoised PET image. Alongside Monte Carlo simulations, experimental results from preclinical and clinical studies showed that the proposed method delivers state-of-the-art denoising performance while retaining spatial resolution and quantification accuracy despite using a common network architecture for various noisy PET images with 1/10th of the full counts. These results highlight the potential of the proposed method to help reduce both PET scan times and the radiation dose of PET tracers substantially without causing additional discomfort to patients.

**Declaration of Competing Interest**

The authors declare that they have no known competing financial interests or personal relationships that could have appeared to influence the work reported in this paper.

**CRediT authorship contribution statement**

**Yuya Onishi:** Conceptualization, Data curation, Formal analysis, Investigation, Methodology, Software, Visualization, Writing - original draft. **Fumio Hashimoto:** Conceptualization, Investigation, Methodology, Resources, Writing - review & editing. **Kibo Ote:** Investigation, Methodology, Resources, Writing - review & editing. **Hiroyuki Ohba:** Investigation, Resources, Writing - review & editing. **Ryosuke Ota:** Investigation, Writing - review & editing. **Etsuji Yoshikawa:** Supervision, Writing - review & editing. **Yasuomi Ouchi:** Supervision, Writing - review & editing.

**Acknowledgments**

The authors would like to thank the scientific advice provided by Dr. Norihiro Harada and Mr. Takashi Isobe from the Central Research Laboratory, Hamamatsu Photonics K.K. We are grateful to Dr. Takeharu Kakiuchi, Dr. Shingo Nishiyama, Mr. Dai Fukumoto, Dr. Masakatsu Kanazawa, Dr. Shigeyuki Yamamoto, and Mr. Takahiro Tsuchiya of the PET Research Group, Central Research Laboratory, Hamamatsu Photonics K.K for providing technical assistance.

**Appendix. Ablation study**

An ablation study was conducted to examine the contributions of each unit in MR-GDD. We denoised PET images by using only the URU or FRU conditional weights or by swapping connections between encoder-decoder and each unit in MR-GDD. To ensure that only certain units worked, we removed a 1 × 1 × 1 3D convolution layer, LReLU, and sigmoid function from the other units. The swap was conducted by connecting the encoder to the FRU and the decoder to the URU, in contrast to normal MR-GDD (see Supplementary Material 0.2.3). Figure A.1 shows the denoised PET images by the ablation study. In addition, the PSNR and SSIM of these images and the line profiles passing through both tumor regions are shown in Table A.1 and Fig. A.2, respectively. These results indicate that the accuracy of the proposed method in terms of the PSNR, SSIM, and contrast recovery was superior than that of the network structures with no unit (deep decoder: DD), a single unit (MR-GDD (Only URU), and MR-GDD (Only FRU)) or the connection the URU and FRU swapped (MR-GDD (Swapped)). The results show that the DD, which did not incorporate the URU and FRU, performed poorly. This shows that it was difficult to generate high-resolution images from random noise using few network parameters without the network structure including the encoder and MR image. Based on the results of the MR-GDD incorporating only the URU, the spatial features from the MR image were strongly weighted and the tumors disappeared. Owing to the lack of an FRU performing a semantic alignment between the output image and the guidance image, a denoised image that emphasized only the spatial details of the MR image was generated. The MR-GDD incorporating only the FRU produced a relatively improved image, compared to DD and MR-GDD (Only URU). However, due to the use of normal upsampling that was unweighted without the URU, a strong bias to promote piecewise smoothness was created, and the small tumor (75th voxel in Fig. A.2) or boundaries (55th voxel in Fig. A.2) in the denoised image were washed away. The URU weights the features derived via upsampling at first, and then the FRU weights each layer between the decoders to function as an effective mechanism. Therefore, the network in which the connection of the URU and FRU was swapped was inferior to the proposed network because the encoder and decoder may possess a useful information for upsampling and semantic feature refinement, respectively. These additional results exhibit similar trends as compared to the attention map results shown in Fig. 5 and Supplementary Material 0.2.4, and we believe that the results of the ablation study emphasize the clear explanation of the proposed method.

Since the MR-DIP network used in this study was a subnetwork of MR-GDD, it had fewer parameters and a lower capacity. It is known that the network structure and capacity can impact the performance of the DIP. Therefore, we evaluated the model to confirm that the performance improvement by MR-GDD was not solely because MR-GDD included more parameters. The network structure remained the same, and the number of blocks (consisting of the combination of a 3 × 3 × 3 3D convolution layer with BN and an LReLU) in the DIP/MR-DIP was increased to match the number of network parameters of MR-GDD. The results are shown in Table A.2. The proposed MR-GDD achieved higher values for PSNR and SSIM even though the total parameters for DIP/MR-DIP (722,337 params) and MR-GDD (709,825 params) were almost identical. Therefore, it is considered that the proposed network structure has a significant influence on the performance improvement of MR-GDD.

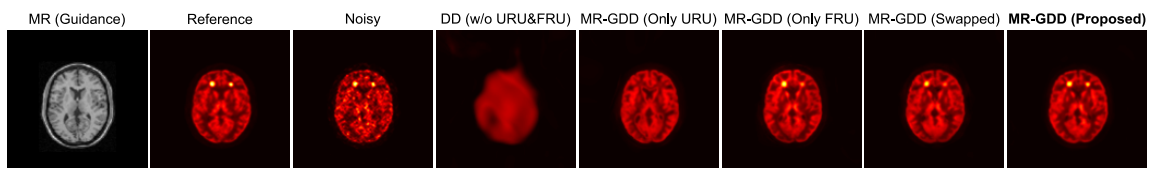

Fig. A.1. Axial images processed by applying different network structures to static brain-PET images simulated using Monte Carlo modeling. From left to right, the columns show sample images of the T1-

weighted MR guidance, reference (150M counts), noisy (5M counts), and the denoised PET images processed using the DD, MR-GDD (Only URU), MR-GDD (Only FRU), MR-GDD (Swapped), and proposed MR-GDD networks.

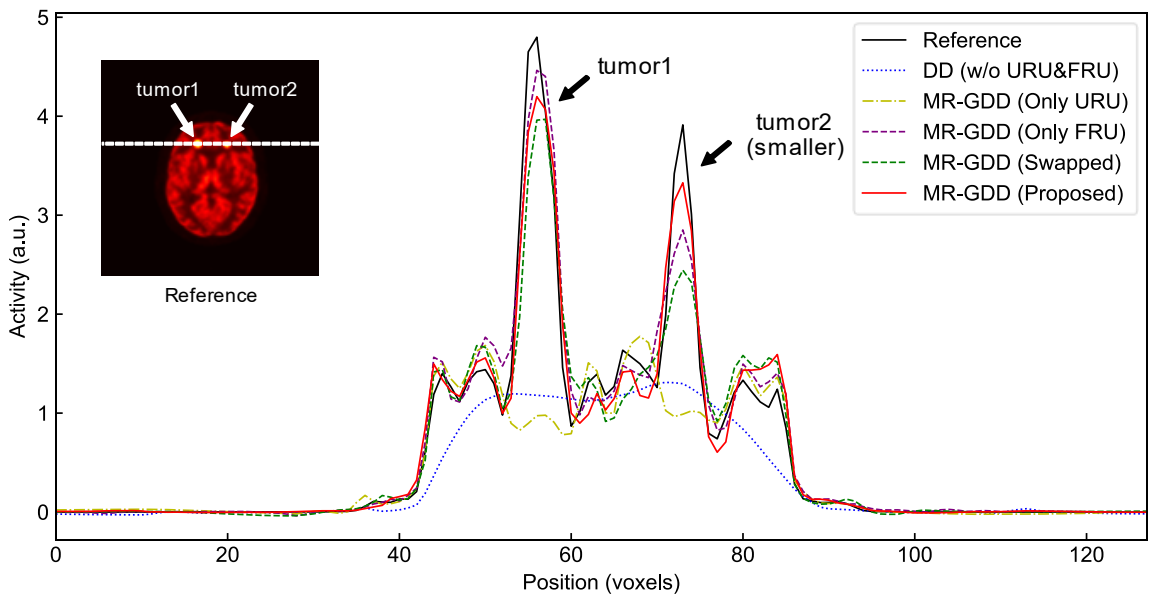

Fig. A.2. Comparison of the line profiles passing through both inserted tumors in the denoised images processed by different network structures. The proposed MR-GDD method retrieved the highest contrast in the area surrounding the smallest inserted tumor (tumor2).

Table A.1

Comparison of the PSNRs and SSIMs for different attention structures in the MR-GDD network. Bold values indicate the best method for each image.

| Method | | URU | FRU | Total params | PSNR [dB] | SSIM |
|---|---|---|---|---|---|---|
| DD | w/o URU & FRU | | | 116,737 | 21.17 | 0.254 |
| MR-GDD | Only URU | ✓ | | 704,545 | 26.72 | 0.835 |
| | Only FRU | | ✓ | 705,601 | 27.68 | 0.883 |
| | Swapped | ✓ | ✓ | 709,825 | 27.69 | 0.880 |
| | Proposed | ✓ | ✓ | 709,825 | **28.33** | **0.886** |

Table A.2

Comparison of the PSNRs and SSIMs with respect to the number of total parameters in each method. Bold values indicate the best method for each image.

| Method | Encoding or Decoding path | Total params | PSNR [dB] | SSIM |
|---|---|---|---|---|
| DIP | 4 blocks | 583,617 | 24.96 | 0.750 |
|        | 5 blocks | 722,337 | 24.65 | 0.689 |
| MR-DIP | 4 blocks | 583,617 | 28.02 | 0.881 |
|        | 5 blocks | 722,337 | 27.65 | 0.877 |
| MR-GDD | 4 blocks | 709,825 | **28.33** | **0.886** |

## References

Arabi, H., Zaidi, H., 2020. Non-local mean denoising using multiple PET reconstructions. Ann. Nucl. Med. 35 (2), 176–186.

Aubert-Broche, B., Griffin, M., Pike, G.B., Evans, A.C., Collins, D.L., 2006. Twenty new digital brain phantoms for creation of validation image data bases. IEEE Trans. Med. Imag. 25 (11), 1410–1416.

Bland, J., Mehranian, A., Belzunce, M.A., Ellis, S., 2018. MR-Guided Kernel EM Reconstruction for Reduced Dose PET Imaging. IEEE Trans. Radiat. Plasma. Med. Sci. 2 (3), 235–243.

Chan, C., Fulton, R., Barnett, R., Feng, D.D., Meikle, S., 2014. Postreconstruction nonlocal means filtering of whole-body PET with an anatomical prior. IEEE Trans. Med. Imag. 33 (3), 635–650.

Chen, K., Gong, E., de Carvalho, F., Xu, J., Boumis, A., Khalighi, M., Poston, K., Sha, S., Greicius, M., Mormino, E., Pauly, J., Srinivas, S., Zaharchuk, G., 2019. Ultra–Low-Dose 18F-Florbetaben Amyloid PET Imaging Using Deep Learning with Multi-Contrast MRI Inputs. Radiology. 290 (3), 649–656.

Çiçek, Ö., Abdulkadir, A., Lienkamp, S.S., Brox, T., Ronneberger, O., 2016. 3D U-Net: learning dense volumetric segmentation from sparse annotation. In Proceedings of the International Conference on Medical Image Computing and Computer-Assisted Intervention, pp. 424–432.

Comtat, C., Kinahan, P.E., Fessler, J.A., Beyer, T., Townsend, D.W., Defrise, M., Michel, C., 2002. Clinically feasible reconstruction of 3D whole-body PET/CT data using blurred anatomical labels. Phys. Med. Biol. 47 (1), 1–20.

Cui, J., Gong, K., Guo, N., Wu, C., Meng, X., Kim, K., Zheng, K., Wu, Z., Fu, L., Xu, B., Zhu, Z., Tian, J., Liu, H., Li, Q., 2019. PET image denoising using unsupervised deep learning. Eur. J. Nucl. Med. Mol. Imaging. 46, 2780–2789.

Fukui, H., Hirakawa, T., Yamashita, T., Fujiyoshi, H., 2019. Attention Branch Network: Learning of Attention Mechanism for Visual Explanation. In Proceedings of the IEEE Conference on Computer Vision and Pattern Recognition, pp. 10705–10714.

Gong, K., Catana, C., Qi, J., Li, Q., 2019. PET Image Reconstruction Using Deep Image Prior. IEEE Trans. Med. Imag. 38 (7), 1655–1665.

Gong, K., Guan, J., Liu, C.-C, Qi, J., 2019. PET Image Denoising Using a Deep Neural Network

Through Fine Tuning. IEEE Trans. Radiat. Plasma Med. Sci. 3 (2), 153–161.

Häggström, I., Schmidtlein, C.R., Campanella, G., Fuchs, T.J., 2019. DeepPET: A deep encoder–decoder network for directly solving the PET image reconstruction inverse problem. Med. Image. Anal. 54, 253–262.

Hamamatsu. Photomultiplier Tubes: Basics and Applications, fourth ed., p. 286. https://www.hamamatsu.com/resources/pdf/etd/PMT_handbook_v4E.pdf.

Hashimoto, F., Ohba, H., Ote, K., Kakimoto, A., Tsukada, H., Ouchi, Y., 2020. 4D Deep Image Prior: Dynamic PET Image Denoising Using an Unsupervised Four-Dimensional Branch Convolutional Neural Network. Phys. Med. Biol. 66 (1), 015006.

Hashimoto, F., Ohba, H., Ote, K., Teramoto, A., Tsukada, H., 2019. Dynamic PET Image Denoising Using Deep Convolutional Neural Networks Without Prior Training Datasets. IEEE Access. 7, 96594–96603.

Hashimoto, F., Ohba, H., Ote, K., Tsukada, H., 2018. Denoising of Dynamic Sinogram by Image Guided Filtering for Positron Emission Tomography. IEEE Trans. Radiat. Plasma Med. Sci. 2 (6), 541–548.

He, K., Sun, J., Tang, X., 2013. Guided Image Filtering. IEEE Trans. Pattern Anal. Mach. Intell. 35 (6), 1397–1409.

Hoifheinz, F., Langner, J., Beuthien-Baumann, B., Oehme, L., Steinbach, J., Kotzerke, J., Hoff, J. 2011. Suitability of bilateral filtering for edge-preserving noise reduction in PET. EJNMMI Res. 1, 23.

ICRP. 2017. Diagnostic reference levels in medical imaging. ICRP Publication 135. Ann. ICRP 46 (1).

Lane, C.A., Parker, T.D., Cash, D.M., Macpherson, K., Donnachie, E., Murray-Smith, H., Barnes, A., Barker S., Beasley, D. G., Bras, J., Brown, D., Burgos, N., Byford, M., Cardoso, M.J., Carvalho, A., Collins, J., De Vita, E., Dickson, J.C., Epie, N., Espak, M., Henley, S.M.D., Hoskote, C., Hutel, M., Klimova, J., Malone, I.B., Markiewicz, P., Melbourne, A., Modat, M., Schrag, A., Shah, S., Sharma, N., Sudre, C.H., Thomas, D.L., Wong, A., Zhang, H., Hardy, J., Zetterberg, H., Ourselin, S., Crutch, S.J., Kuh, D., Richards, M., Fox, N.C., Schott, J.M., 2017. Study protocol: Insight 46 – a neuroscience sub-study of the MRC National Survey of Health and Development. BMC Neurol. 17 (1), 75.

Lehtinen, J., Munkberg, J., Hasselgren, J., Laine, S., Karras, T., Aittala, M., Aila, T., 2018. Noise2Noise: Learning Image Restoration without Clean Data. Proc. Mach. Learn. Res. 80, 2965–2974.

Lin, Y.C., Huang, H.M., 2020. Denoising of multi b-value diffusion-weighted MR images using deep image prior. Phys. Med. Biol. 65, 105003.

Litjens, G., Kooi, T., Bejnordi, B.E., Setio, A.A.A, Ciompi, F., Ghafoorian, M., van der Laak,

J.A.W.M, van Ginneken B., Sánchez, C.I., 2017. A survey on deep learning in medical image analysis. Med. Image. Anal. 42, 60–88.

Liu, C.-C., Qi, J., 2019. Higher SNR PET image prediction using a deep learning model and MRI image. Phys. Med. Biol. 64 (11), 115004.

Markiewicz, P.J., Cash, D., Schott, J.M., 2018. Single amyloid PET scan on the Siemens Biograph mMR. Zenodo. https://doi.org/10.5281/ZENODO.1472951.

Markiewicz, P.J., Ehrhardt, M.J., Erlandsson, K., Noonan, P.J., Barnes, A., Schott, J.M., Atkinson, D., Arridge, S.R., Hutton, B.F., Ourselin, S., 2018. NiftyPET: a High-throughput Software Platform for High Quantitative Accuracy and Precision PET Imaging and Analysis. Neuroinformatics. 16, 95–115.

Ote, K., Hashimoto, F., Kakimoto, A., Isobe, T., Inubushi, T., Ota, R., Tokui, A., Saito, A., Moriya, T., Omura, T., Yoshikawa, E., Teramoto, A., Ouchi, Y., 2020. Kinetics-Induced Block Matching and 5D Transform Domain Filtering for Dynamic PET Image Denoising. IEEE Trans. Radiat. Plasma Med. Sci. 4 (6), 720–728.

Paxinos, G., Huang, X.F., Toga, A.W., 2000. The Rhesus Monkey Brain in Stereotaxic Coordinates. Academic Press. ISBN 0-12-358255-5.

Phelps, M.E., 2012. PET: Molecular Imaging and Its Biological Applications. Springer, New York.

Sanaat, A., Arabi, H., Mainta, I., Garibotto, V., Zaidi, H., 2020. Projection Space Implementation of Deep Learning–Guided Low-Dose Brain PET Imaging Improves Performance over Implementation in Image Space. J. Nucl. Med. 61(9), 1388–1396.

Saul, L.K., Roweis, S.T., 2003. Think globally, fit locally: unsupervised learning of low dimensional manifolds. J. Mach. Learn. Res. 4. 119–155.

Schlemper, J., Oktay, O., Schaap, M., Heinrich, M., Kainz, B., Glocker, B., Rueckert, D., 2019. Attention gated networks: Learning to leverage salient regions in medical images. Med. Image. Anal. 53, 197–207.

Schramm G., Rigie, D., Vahle, T., Rezaei, A., Van Laere, K., Shepherd, T., Nuyt, J., Boada, F., 2021. Approximating anatomically-guided PET reconstruction in image space using a convolutional neural network. NeuroImage. 224 (1), 117399.

Spuhler, K., Serrano-Sosa, M., Cattell, R., DeLorenzo, C., Huang, C., 2020. Full-count PET recovery from low-count image using a dilated convolutional neural network. Med. Phys. 47 (10), 4928–4938.

Sudarshan, V.P., Egan, G.F., Chen, Z., Awate, S.P., 2020. Joint PET-MRI image reconstruction using a patch-based joint-dictionary prior. Med. Image. Anal. 62, 101669.

Tachella, J., Tang, J., Davies, M., 2020. The Neural Tangent Link Between CNN Denoisers and Non-Local Filters. arXiv. 2006.02379.

Tanaka, E., Kudo, H., 2010. Optimal Relaxation Parameters of DRAMA (Dynamic RAMLA) Aiming

at One-pass Image Reconstruction for 3D-PET. Phys. Med. Biol. 55 (10), 2917–2939.

Tashima, H., Yoshida, E., Iwao, Y., Wakizaka, H., Maeda, T., Seki, C., Kimura, Y., Takado, Y., Higuchi, M., Suhara, T., Yamashita, T., Yamaya, T., 2019. First prototyping of a dedicated PET system with the hemi-sphere detector arrangement. Phys. Med. Biol. 64 (6), 065004.

Uezato, T., Hong, D., Yokoya, N., He, W., 2020. Guided Deep Decoder: Unsupervised Image Pair Fusion. In Proceedings of the European Conference on Computer Vision, pp. 87–102.

Ulyanov, D., Vedaldi, A., Lempitsky, V., 2018. Deep image prior. In Proceedings of the IEEE Conference on Computer Vision and Pattern Recognition, pp. 9446–9454.

Vunckx, K., Are, A., Baete, K., Reihac, A., Deroose, C.M., Van Laere, K., Nuyts, J., 2012. Evaluation of three MRI-based anatomical priors for quantitative PET brain imaging. IEEE Trans. Med. Imag. 31 (3), 599–612.

Watanabe, M., Saito, A., Isobe, T., Ote, K., Yamada, R., Moriya, T., Omura, T., 2017. Performance evaluation of a high-resolution brain PET scanner using four-layer MPPC DOI detectors. Phys. Med. Biol. 62 (17), 7148–7166.

Wong, D.F., Rosenberg, P.B., Zhou, Y., Kumar, A., Raymont, V., Ravert, H.T., Dannals, R.F., Nandi, A., Brašić, J.R., Ye, W., Hilton, J., Lyketsos, C., Kung, H.F., Joshi, A.D., Skovronsky, D.M., Pontecorvo, M.J., 2010. In vivo imaging of amyloid deposition in Alzheimer disease using the radioligand $^{18}$F-AV-45 (Florbetapir F 18). J. Nucl. Med. 51 (6), 913–920.

Xu, Z., Gao, M., Papadakis, G.Z., Luna, B., Jain, S., Mollura, D.J., Bagci, U., 2018. Joint solution for PET image segmentation, denoising, and partial volume correction. Med. Image. Anal. 46, 229–243.

Yan, J., Lim, J.C.-S., Townsend, D.W., 2015. MRI-guided brain PET image filtering and partial volume correction. Phys. Med. Biol. 60 (3), 961–976.

Yokota, T., Hontani, H., Zhao, Q., Cichocki, A., 2020. Manifold Modeling in Embedded Space: An Interpretable Alternative to Deep Image Prior. IEEE Trans. Neural Netw. Learn. Syst. doi: 10.1109/TNNLS.2020.3037923.

Zhou, L., Schaefferkoetter, J.D., Tham, I.W.K., Hung, G., Yan, J., 2020. Supervised learning with cyclegan for low-dose FDG PET image denoising. Med. Image. Anal. 65, 101770.

Zhu, C., Byrd, R.H., Lu, P., Nocedal, J., 1997. Algorithm 778: L-BFGS-B: Fortran subroutines for large-scale bound-constrained optimization. ACM Trans. Math. Softw. 23 (4), 550–560.